\documentclass[aps,notitlepage,onecolumn,preprintnumbers,superscriptaddress,amsmath,amssymb,showpacs,nofootinbib]{revtex4-1}
\usepackage{graphicx}% Include figure files
\usepackage{dcolumn}% Align table columns on decimal point
\usepackage{bm}% bold math
\usepackage{hyperref}% add hypertext capabilities
\hypersetup{
	colorlinks=true,
	linkcolor=blue,
	citecolor=blue,
	urlcolor=blue
	}
\usepackage[mathlines]{lineno}% Enable numbering of text and display math
\usepackage{color}
\usepackage{xcolor}

\begin{document}

\title{Investigating the Isospin Property of $T_{cc}^+$ from its Dalitz Plot Distribution}

\author{Jun Shi} 
\email{\text{jun.shi@scnu.edu.cn}}
\author{Enke Wang} 
\email{\text{wangek@scnu.edu.cn}}
\author{Qian Wang}
\email{\text{qianwang@m.scnu.edu.cn}}
\affiliation{Guangdong Provincial Key Laboratory of Nuclear Science, Institute of Quantum Matter, South
China Normal University, Guangzhou 510006, China }
\affiliation{Guangdong-Hong Kong Joint Laboratory of Quantum Matter, Southern Nuclear Science Computing
Center, South China Normal University, Guangzhou 510006, China }

%\date{\today}
%%%%%%%%%%%%%%%%%%%%%%%%%%%%%%%%%%%%%%%%%%%%%%%%%%%%%%%%%%%%%%%%%%%%%%%%%%%%%

\begin{abstract} 
Recently, LHCb observed a double-charmed tetraquark candidate $T_{cc}^+$ and claimed its isospin to be zero due to the absence of $T_{cc}^{++}$. However, the absence of the $T_{cc}^{++}$ in LHCb may be due to their low production in $pp$ collisions, thus the isospin of $T_{cc}^+$ still needs to be clarified. We propose an efficient way to investigate the isospin property of $T_{cc}^+$ from the Dalitz plot distribution of $T_{cc}^+\to D^+D^0\pi^0$. This method can also be applied to other exotic candidates.
\end{abstract} 

\maketitle

%%%%%%%%%%%%%%%%%%%%%%%%%%%%%%%%%%%%%%%%%%%%%%%%%%%%%%%%%%%%%%%%%%%%%%%%%%%%%

\section{Introduction}
Exploring the inner structure of hadrons and their dynamical properties has been one of the major
targets of particle physics. The conventional quark model, in terms of constituent quark degrees of freedom,
classifies hadrons into mesons made of quark and antiquark and baryons
constructed from three quarks. It can explain the $S$-wave ground hadrons and successfully 
predict the existence of the $\Omega$ baryon. In the same degrees of freedom,
the non-relativistic quark model can also describe the spectra of heavy quarkonium
below open-flavor threshold. Above open-flavor threshold, the experimental measurements
are not consistent with the expectations of non-relativistic quark model. 
For instance, the famous $X(3872)$~\cite{Belle:2003nnu}, $Z_c(3900)$~\cite{BESIII:2013ris,Belle:2013yex,Xiao:2013iha}, $Z_c(4020)$~\cite{BESIII:2013ouc}, $T_{cc}^+$~\cite{LHCb:2021vvq,LHCb:2021auc},
$X(6900)$~\cite{LHCb:2020bwg} and so on. To understand their nature, numerous efforts have been put forward.
For reviews of these particles, we would recommend Refs.~\cite{Chen:2016qju,Liu:2019zoy,Chen:2016spr,Dong:2017gaw,Lebed:2016hpi,Guo:2017jvc,Albuquerque:2018jkn,Yamaguchi:2019vea,Guo:2019twa,Brambilla:2019esw}.
For the properties of these exotic candidates, most of them
are accepted by either hadronic molecular or compact multi-quark pictures. In addition, for exotic candidates
with quantum numbers allowed by conventional quark model, they could also be accepted as
conventional hadrons with a mixture of nearby continuum, for instance the $X(3872)$ as 
normal $\chi_{c1}(2P)$ charmonium with a mixture of $D\bar{D}^*+c.c.$ component~\cite{Meng:2005er}.

Among them, the $T_{cc}^+$ is the first doubly charmed exotic candidate
discovered by LHCb ~\cite{LHCb:2021vvq,LHCb:2021auc}
in $D^0D^0\pi^+$ channel, which arouse
high research enthusiasm~\cite{Hudspith:2020tdf,Cheng:2020wxa,Qin:2020zlg,Pflaumer:2020ogv,Drutskoy:2021euy,Agaev:2021vur,Dong:2021bvy,Weng:2021hje,Ren:2021dsi,Du:2021zzh,Aliev:2021dgx,Ozdem:2021hmk,Agaev:2022ast,Kim:2022mpa,He:2022rta,Ozdem:2022yhi,Meng:2022ozq,Hu:2021gdg}.  Because it locates close to the $DD^*$
threshold, it could be considered as a $DD^*$ hadronic molecule.
On the other hand, based on its observed channel,
it contains constituent $cc\bar{u}\bar{d}$ and could also be viewed as
compact $cc\bar{u}\bar{d}$ tetraquark. Even before its observation,
intensive discussions about whether this kind of state exists~\cite{Pepin:1996id,Gelman:2002wf,Vijande:2003ki,Janc:2004qn,Yang:2009zzp,Lee:2009rt,Ebert:2007rn,Vijande:2007rf,Navarra:2007yw,Li:2012ss,Feng:2013kea,Ikeda:2013vwa,Karliner:2017qjm,Wang:2017uld,Junnarkar:2018twb,Deng:2018kly,Liu:2020nil} or not~\cite{Semay:1994ht,Janc:2004qn,Yang:2009zzp,Eichten:2017ffp,Park:2018wjk,Lu:2020rog,Cheng:2020wxa,Braaten:2020nwp,Meng:2021jnw}
as well as its mass position if exists.  In the molecular picture,
the $T_{cc}^+$ dominantly couples to the $D^{\ast+}D^0$ and
$D^{\ast0}D^+$ channels. For the $S$-wave interaction, the favored quantum number is $J^P=1^+$ with
isospin not well determined, either to be 0 or 1. As the result,
the observed $T_{cc}^+$ could be either $T_{cc}^+(I=0,~I_z=0)$ or $T_{cc}^{\prime +}(I=1,~I_z=0)$
or a mixture of them. Here we denote isospin singlet and triplet as $T_{cc}$ and $T_{cc}^{\prime}$~\cite{Hu:2021gdg}, respectively.
If it is isospin triplet, the other two partners, 
i.e. $T_{cc}^{\prime ++}$ and $T_{cc}^{\prime 0}$ should be present.
However, $T_{cc}^{\prime++}$ does not show
up in the observation of LHCb, and for this reason LHCb claimed the isospin of $T_{cc}^+$ should be 0.
Very recently, Hu et al.~\cite{Hu:2021gdg} found that comparing with the central collisions, the
yields of $T_{cc}$ is strongly suppressed, about three orders of magnitude, in the very peripheral collisions,
say $pp$ collision, which is the case of LHCb
experiment~\cite{LHCb:2021vvq,LHCb:2021auc}. This indicates that
the absence of $T_{cc}^{\prime ++}$ and $T_{cc}^{\prime 0}$ in LHCb
experiment~\cite{LHCb:2021vvq,LHCb:2021auc} is probably due to
their small production rate in $pp$ collision. Their existence needs future experiments to verify. Therefore,
the isospin of $T_{cc}^+$ needs further investigation.

In this work, we propose an efficient way to distinguish the isospin of $T_{cc}^{+}$ from the Dalitz
plot distribution of $T_{cc}^{+}\to D^+D^0\pi^0$.  Sec.~\ref{sec:theory}
presents our framework. The results and discussions follow as Sec.~\ref{sec:method}.
Sec.~\ref{sec:summary} is at the end.

\section{Theoretical Formalism}\label{sec:theory}
The scattering between an isospin $\frac 12$ doublet $\mathcal{D}=(D^+,D^0)$ 
and an isospin $\frac 12$ doublet $\mathcal{D}^*=(D^{*+},D^{*0})$ could form an isospin singlet molecule 
\begin{equation}
|T_{cc}^+\rangle = -\frac{1}{\sqrt{2}}(|D^{\ast+}\rangle|D^{0}\rangle - |D^{\ast 0}\rangle|D^{+}\rangle),\quad I=0\quad I_z=0
\end{equation}
and an isospin triplet molecules
\begin{eqnarray}
|T_{cc}^{\prime ++}\rangle &=& |D^{\ast+}\rangle|D^{+}\rangle,\quad I=1\quad I_z=+1,\\
|T_{cc}^{\prime+}\rangle &=& -\frac{1}{\sqrt{2}}(|D^{\ast+}\rangle|D^{0}\rangle + |D^{\ast 0}\rangle|D^{+}\rangle),\quad I=1\quad I_z=0,\\
|T_{cc}^{\prime 0}\rangle &=& |D^{\ast 0}\rangle|D^{0}\rangle,\quad I=1\quad I_z=-1.
\end{eqnarray}
The isospin information of doubly charmed exotic candidate $T_{cc}^+$, up to now, cannot be well determined
from experimental side. The LHCb collaboration claims that they do not find signal of 
the isospin partner $T_{cc}^{\prime ++}$, which indicates the observed $T_{cc}^+$ 
is an isospin singlet doubl charmed tetraquark. The absence of the $T_{cc}^{\prime ++}$
in $pp$ collision may because that either the potential between $D^{*+}D^+$ is
repulsive and cannot be form a bound state or the production of the isospin triplet state
is suppressed. If the latter one is the case, one would expect to observed the isospin triplet
partners in other reactions, for instance in central collision of heavy ion collision~\cite{Hu:2021gdg}.  
In this case, one cannot avoid a mixing between the $T_{cc}^+$ and the $T_{cc}^{\prime +}$.
Thus an efficient way to investigate the isospin property is timely.  

\begin{figure}[htbp]
%\hspace*{-0.5cm}
\includegraphics[width=0.4\textwidth]{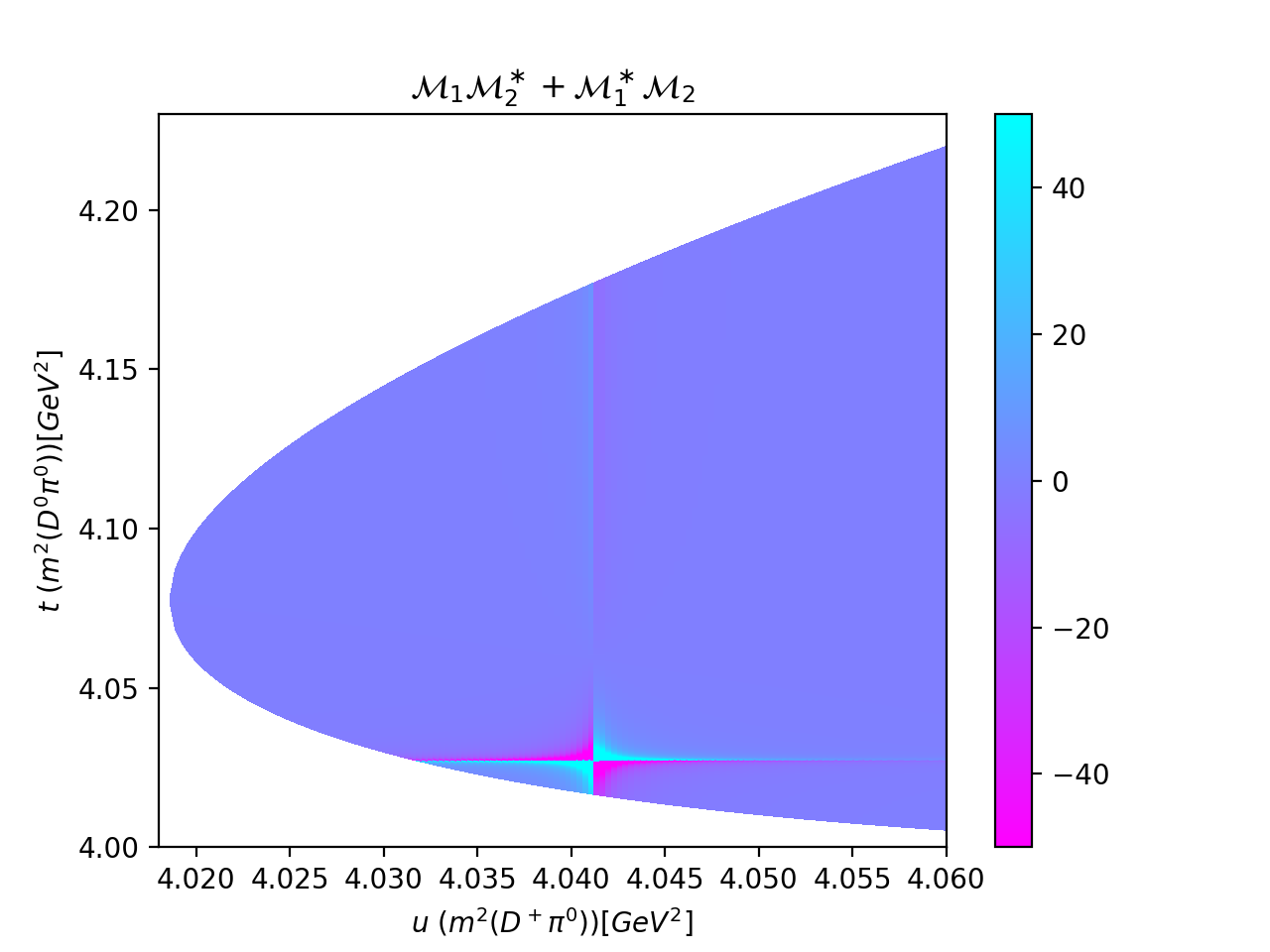}\hspace{0.5cm} \includegraphics[width=0.4\textwidth]{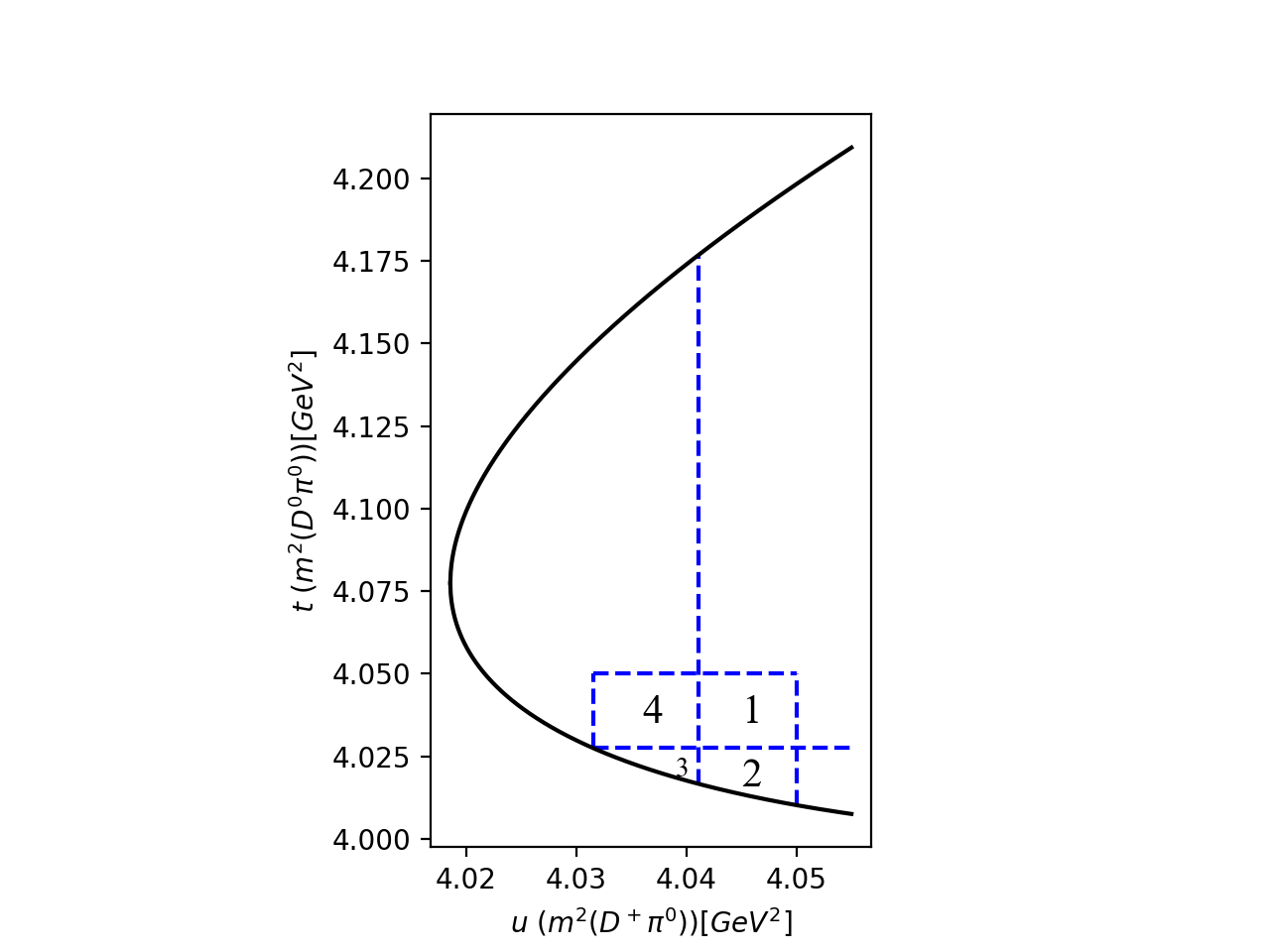}
\caption{The interference term shows significant difference around $t_a = m^2_{D^{\ast 0}}$, $u_a = m^2_{D^{\ast +}}$ (left panel) in Dalitz plot. The right panel shows the separation divided by $t_a = m^2_{D^{\ast 0}}$, $t_b=4.05~\mathrm{GeV}^2$,
$u_b = 4.032 ~\mathrm{GeV}^2$, $u_a = m^2_{D^{\ast +}}$, and $u_c =4.05~\mathrm{GeV}^2$, where $u_b$ is the $u$-value of the intersection of $t_a = m^2_{D^{\ast 0}}$ and the
left-bottom-border of the Dalitz plot.
The four regions are labeled as ``1", ``2", ``3", and ``4", respectively.}
\label{fig:interference}
\end{figure}

Searching for pure isospin triplet state, for instance the $T_{cc}^{\prime ++}$
is a way to pin down the isospin property of $T_{cc}^+$. On the other hand, the interference could also give an 
efficient way to explore its isospin property. Thus we propose to study the isospin 
information in the $D^+D^0\pi^0$ channel, as it can occur via the two 
intermediated states $D^{\ast+}$ and $D^{\ast+}$, i.e. $T_{cc}^+\to D^{\ast+}D^{0}\to D^+D^0\pi^0 $ and $T_{cc}^+\to D^{\ast 0}D^{+}\to D^+D^0\pi^0 $.  Dalitz plot would be a prominent measurement to discover
the isospin of $T_{cc}^+$. Firstly, we construct Lagrangian for each interaction 
in the process. 
The charm-light meson in SU(2) flavor symmetry reads as
\[
\mathcal{D}^{(*)}=\left(\begin{array}{c}
D^{(*)0}(c\bar{u})\\
D^{(*)+}(c\bar{d}))
\end{array}\right)\quad\bar{\mathcal{D}}^{(*)}=\left(\begin{array}{c}
\bar{D}^{(*)0}(\bar{c}u)\\
D^{(*)-}(\bar{c}d)
\end{array}\right).
\]
In the heavy quark limit, the $S$-wave ground $D$ and $D^*$
mesons form a doublet, with the superfield
\[
H_{a}=P_{a}+V\cdot\sigma
\]
which annihilates $Q\bar{q}$ meson field,
and its anti-superfield
\[
\bar{H}_{a}=\bar{P}_{a}-\bar{V}_{a}\cdot\sigma
\]
which annihilates $\bar{Q}q$ meson field.
They behave as
\begin{align*}
H_{a} & \xrightarrow{\mathcal{P}}-H_{a},\quad H_{a}\xrightarrow{\mathcal{C}}\sigma_{2}\bar{H}_{a}^{T}\sigma_{2},\quad H_{a}\xrightarrow{\mathcal{S}}SH_{a},\quad H_{a}\xrightarrow{\mathcal{U}}H_{b}U_{ba}^{\dagger}\\
\bar{H}_{a} & \xrightarrow{\mathcal{P}}-\bar{H}_{a},\quad\bar{H}_{a}\xrightarrow{\mathcal{C}}\sigma_{2}H_{a}^{T}\sigma_{2},\quad\bar{H}_{a}\xrightarrow{\mathcal{S}}\bar{H}\bar{S}^{\dagger},\quad\bar{H}_{a}\xrightarrow{\mathcal{U}}U_{ab}\bar{H}_{b}
\end{align*}
under parity $\mathcal{P}$, charge parity $\mathcal{C}$, heavy quark spin $\mathcal{S}$, and chiral $\mathcal{U}$ transformations. Their pion coupling reads as
\begin{align*}
\mathcal{L}_{\pi} & =\frac{g_{Q}}{4}\langle\sigma\cdot u_{ab}\bar{H}_{b}\bar{H}_{a}^{\dagger}\rangle+h.c.\\
 & =\frac{g_{Q}}{4}\langle\sigma^{i}(-\nabla^{i}\tau_{ab}^{\alpha}\phi^{\alpha})\bar{H}_{b}\bar{H}_{a}^{\dagger}\rangle+h.c.,
\end{align*}
with 
\begin{align*}
u & =-\nabla\Phi/f_{\pi}+\mathcal{O}(\Phi^{3})\\
 & =-\text{\ensuremath{\nabla\tau^{a}\phi^{a}/f_{\pi}+\mathcal{O}(\Phi^{3})}},
\end{align*}
\[
\Phi=\left(\begin{array}{cc}
\pi^{0} & \sqrt{2}\pi^{+} \\
\sqrt{2}\pi^{-} & -\pi^{0} \\
\end{array}\right),
\]
and $f_{\pi}=92.4$MeV. Expanding explicitly, one can obtain
\begin{align*}
\mathcal{L}_{\pi}^{(1)} & =\frac{g_{Q}}{2f_{\pi}}\left(i\epsilon^{ijk}V_{a}^{j\dagger}V_{b}^{k}+V_{a}^{i\dagger}P_{b}+P_{a}^{\dagger}V_{b}^{i}\right)\partial^{i}\Phi_{ba}\\
 & +\frac{g_{Q}}{2f_{\pi}}\left(i\epsilon^{ijk}\bar{V}_{a}^{j\dagger}\bar{V}_{b}^{k}+\bar{P}_{a}^{\dagger}V_{b}^{i}+\bar{V}_{a}^{i\dagger}\bar{P}_{b}\right)\partial^{i}\Phi_{ab}
\end{align*}
with $g_{Q}=0.57.$ Notice that the factor $\sqrt{2m}$ is implicit
for each heavy field with $m$ the mass of the corresponding heavy
field. Accordingly, the relevant partial width of $D^{*0}$ and $D^{*+}$
read as
\begin{align*}
\Gamma(D^{*+} & \to D^{0}\pi^{+})=\frac{g_{Q}^{2}m_{D^{0}}k_{\pi^{+}}^{3}}{12\pi f_{\pi}^{2}m_{D^{*+}}}\\
\Gamma(D^{*+} & \to D^{+}\pi^{0})=\frac{g_{Q}^{2}m_{D^{+}}k_{\pi^{0}}^{3}}{24\pi f_{\pi}^{2}m_{D^{*+}}}\\
\Gamma(D^{*0} & \to D^{+}\pi^{-})=\frac{g_{Q}^{2}m_{D^{+}}k_{\pi^{-}}^{3}}{12\pi f_{\pi}^{2}m_{D^{*+}}}\\
\Gamma(D^{*0} & \to D^{0}\pi^{0})=\frac{g_{Q}^{2}m_{D^{0}}k_{\pi^{0}}^{3}}{24\pi f_{\pi}^{2}m_{D^{*+}}}
\end{align*}
with $k_{\pi}$ the three-momentum of pion in the rest frame of the
decaying charm-light meson. Here as we want to extract the isospin
information of $T_{cc}$, the mass difference between charged and
neutral particles has been considered. The denominator of the propagator of vector $\mathcal{D}^{*}$
meson works in the Flatte form 
\begin{align*}
\mathcal{D}_{D^{*+}}(E) & =E-E_{f}+\frac{i}{2}\Gamma(D^{*+}\to D^{0}\pi^{+})+\frac{i}{2}\Gamma(D^{*+}\to D^{+}\pi^{0}),\\
\mathcal{D}_{D^{*0}}(E) & =E-E_{f}+\frac{i}{2}\Gamma(D^{*0}\to D^{0}\pi^{0})+\frac{i}{2}\Gamma(D^{*0}\to D^{+}\pi^{-}).
\end{align*}
Here 
\[
k_{\pi}=\sqrt{2\mu(E-m_{a}-m_{b})}\Theta(E-m_{a}-m_{b})+i\sqrt{2\mu(m_{a}+m_{b}-E)}\Theta(m_{a}+m_{b}-E)
\]
$m_{a}$ and $m_{b}$ the masses of the final particles in the considered
channel. The parameter $E_{f}$ is related to bare pole $m_{0}$ and
finally gives the right pole position of the physical $D^{*+}$ and
$D^{*0}.$ In total, the amplitudes of $T_{cc}^{+}\to D^{*+}D^{0}\to D^{+}\pi^{0}D^{0}$
and $T_{cc}^{+}\to D^{*0}D^{+}\to D^{+}\pi^{0}D^{0}$ are
\begin{align*}
\mathcal{M}_1\equiv\mathcal{M}(D^{*0}) & =\frac{g_{D^{*0}D^{+}T_{cc}}\frac{g_{Q}}{2f_{\pi}}\sqrt{2m_{D^{*0}}2m_{D^{0}}}k_{\pi^{0}}}{\mathcal{D}_{D^{*0}}(E_{D^{0}\pi^{0}})},\\
\mathcal{M}_2\equiv\mathcal{M}(D^{*+}) & =\frac{g_{D^{*+}D^{0}T_{cc}}\frac{g_{Q}}{2f_{\pi}}\sqrt{2m_{D^{*+}}2m_{D^{+}}}k_{\pi^{0}}}{\mathcal{D}_{D^{*+}}(E_{D^{+}\pi^{0}})},
\end{align*}
respectively. Here $k_{\pi^{0}}$ equal to zero when $E$ is smaller than the relevant
threshold.  The relative ratio between the couplings
\[
r\equiv \frac{g_{D^{*+}D^{0}T_{cc}}}{g_{D^{*0}D^{+}T_{cc}}}
\]
depends on the isospin property of $T_{cc}$, 
i.e. $r=-1$ and $r=1$ for isospin singlet and triplet, respectively.
In the following, one will see that only the relative ratio $r$ is the key value to determine the isospin property of the 
$T_{cc}^+$. Supposing the physical $T_{cc}^+$ state is a combination of isospin 0 and 1 states, one can express the
decaying amplitude of
$T_{cc}^+\to D^+D^0\pi^0$ as
\begin{equation}
\mathcal{M} \equiv \frac{1}{\sqrt{1+r^2}}(\mathcal{M}_1 + r \mathcal{M}_2),
\end{equation}
with $r\in [-1,~1]$, and ${1}/{\sqrt{1+r^2}}$ is a
normalization factor. For the one to three decay process $T_{cc}^+\to D^+D^0\pi^0$,
one can define Mandelstam variables as
$t \equiv (p_{D^0} + p_{\pi^0})^2$, $u \equiv (p_{D^+} + p_{\pi^0})^2$, and
$s \equiv (p_{D^0} + p_{D^+})^2$. They are not independent as 
\begin{equation}
s+t+u=M_{T_{cc}^+}^2+M_{D^{+}}^2+M_{D^0}^2+M_{\pi^0}^2.
\end{equation}
As there are resonance contributions in the $D^+\pi^0$ and $D^0\pi^0$
subsystems, it is easy to use the Mandelstam variables $t$ and $u$ 
as two independent variables in the following analysis. 
The Dalitz plot distribution of $T_{cc}^+\to D^+D^0\pi^0$ in terms of $t$ and
$u$ depends on the absolute amplitude square, which is
\begin{equation}
	|\mathcal{M}|^2 = \frac{1}{1+r^2}\left[|\mathcal{M}_1|^2 + r^2 |\mathcal{M}_2|^2 +
	r\left(\mathcal{M}_1\mathcal{M}_2^\ast + \mathcal{M}_1^\ast \mathcal{M}_2\right)\right].\label{eq:amp_square_r}
	\end{equation}
Our goal is to define an observable quantity
from a special section of the Dalitz plot, which has a monotonic dependence on linear $r$. Thus a certain
value of the quantity will have a one-to-one correspondence with $r$, which can tell us the isospin property
of $T_{cc}^+$. The last two terms of Eq.~\eqref{eq:amp_square_r} is the 
key for the isospin property, which exhibits significant difference around
$t_a\equiv M_{D^{*0}}^2$ and  $u_a\equiv M_{D^{*+}}^2$ as shown by the left 
panel of Fig.~\ref{fig:interference}.

Supposing we define the quantity proportional to be the subtraction of events of area 1 (noted as $N_{A_1}$) and
that of area 2 (noted as $N_{A_2}$), which is
\begin{equation}
	A'=\frac{N_{A_1} - N_{A_2}}{N_{A_1} + N_{A_2}}=\frac{\int_{A_1-A_2}|\mathcal{M}_1|^2dtdu +r^2\int_{A_1-A_2}|\mathcal{M}_2|^2dtdu
	+r\int_{A_1-A_2}\left(\mathcal{M}_1\mathcal{M}_2^\ast + \mathcal{M}_1^\ast \mathcal{M}_2\right)dtdu
	}{
	\int_{A_1+A_2}|\mathcal{M}_1|^2dtdu +r^2\int_{A_1+A_2}|\mathcal{M}_2|^2dtdu
	+r\int_{A_1+A_2}\left(\mathcal{M}_1\mathcal{M}_2^\ast + \mathcal{M}_1^\ast \mathcal{M}_2\right)dtdu
	}\label{eq:test_quantity}
\end{equation}
where
\begin{eqnarray}
	\int_{A_1 \pm A_2}|\mathcal{M}_1|^2dtdu &=& \int_{A_1}|\mathcal{M}_1|^2dtdu
	\pm \int_{A_2}|\mathcal{M}_1|^2dtdu,\\
	\int_{A_1 \pm A_2}|\mathcal{M}_2|^2dtdu &=& \int_{A_1}|\mathcal{M}_2|^2dtdu
	\pm \int_{A_2}|\mathcal{M}_2|^2dtdu,\\
   \int_{A_1 \pm A_2}\left(\mathcal{M}_1\mathcal{M}_2^\ast + \mathcal{M}_1^\ast \mathcal{M}_2\right)dtdu &=& \int_{A_1}\left(\mathcal{M}_1\mathcal{M}_2^\ast + \mathcal{M}_1^\ast \mathcal{M}_2\right)dtdu
   \pm \int_{A_2}\left(\mathcal{M}_1\mathcal{M}_2^\ast + \mathcal{M}_1^\ast \mathcal{M}_2\right)dtdu.
\end{eqnarray}
From Eq. (\ref{eq:test_quantity}), we can see that the denominator and numerator are qudratic functions of
$r$, and the behavior of the quantity in terms of $r$ depends on the subtraction of the events in $A_1$ and
$A_2$. 
To get a monotonic
behavior for the quantity with the variance of $r$, it is crucial to choose the areas properly
to eliminate the contributions of terms multiplying $r^2$, which contains $|\mathcal{M}_2|^2$, and make the terms
multiplying $r$ dominant, which is the interference term. Because of the propagators of $D^{*0}$ and $D^{*+}$ in $\mathcal{M}_1$ and $\mathcal{M}_2$, the Dalitz plot distribution has great
enhancements along $t_a \equiv m^2_{D^{\ast 0}}$ and $u_a \equiv m^2_{D^{\ast +}}$. The interference term get most of its biggest values around the intersection of $t_a$ and $u_a$, as shown by the left panel of Fig. \ref{fig:interference}.
We find an essential
way to define the quantity, which can reduce the contribution of $|\mathcal{M}_2|^2$ relating to the quadratic term of $r$, and highlight that of the interference term, i.e. the linear term of $r$.
The quantity is defined as
\begin{equation}
A \equiv \frac{N_1 + N_3 - N_2 - N_4}{N_1 + N_3 + N_2 + N_4},
\label{eq:ratio}
\end{equation}
with the partition explained by the right panel of Fig. \ref{fig:interference}. 
Here $N_i$ is the number of events in the $i$th region.
The values of $t_a = m^2_{D^{\ast 0}}$, $u_a = m^2_{D^{\ast +}}$, $t_b=4.05~\mathrm{GeV}^2$,
$u_b = 4.05 ~\mathrm{GeV}^2$ and $u_c =4.03~\mathrm{GeV}^2$ are chosen to give a typical monotonic dependence of
$A$ on $r$, changing those values slightly, say of order $0.001~\mathrm{GeV}^2$, will not affect the result significantly. Larger area in real experimental observation could involve in the effects of background.
\begin{figure}[htbp]
	%\hspace*{-0.5cm}
	\includegraphics[width=0.8\textwidth]{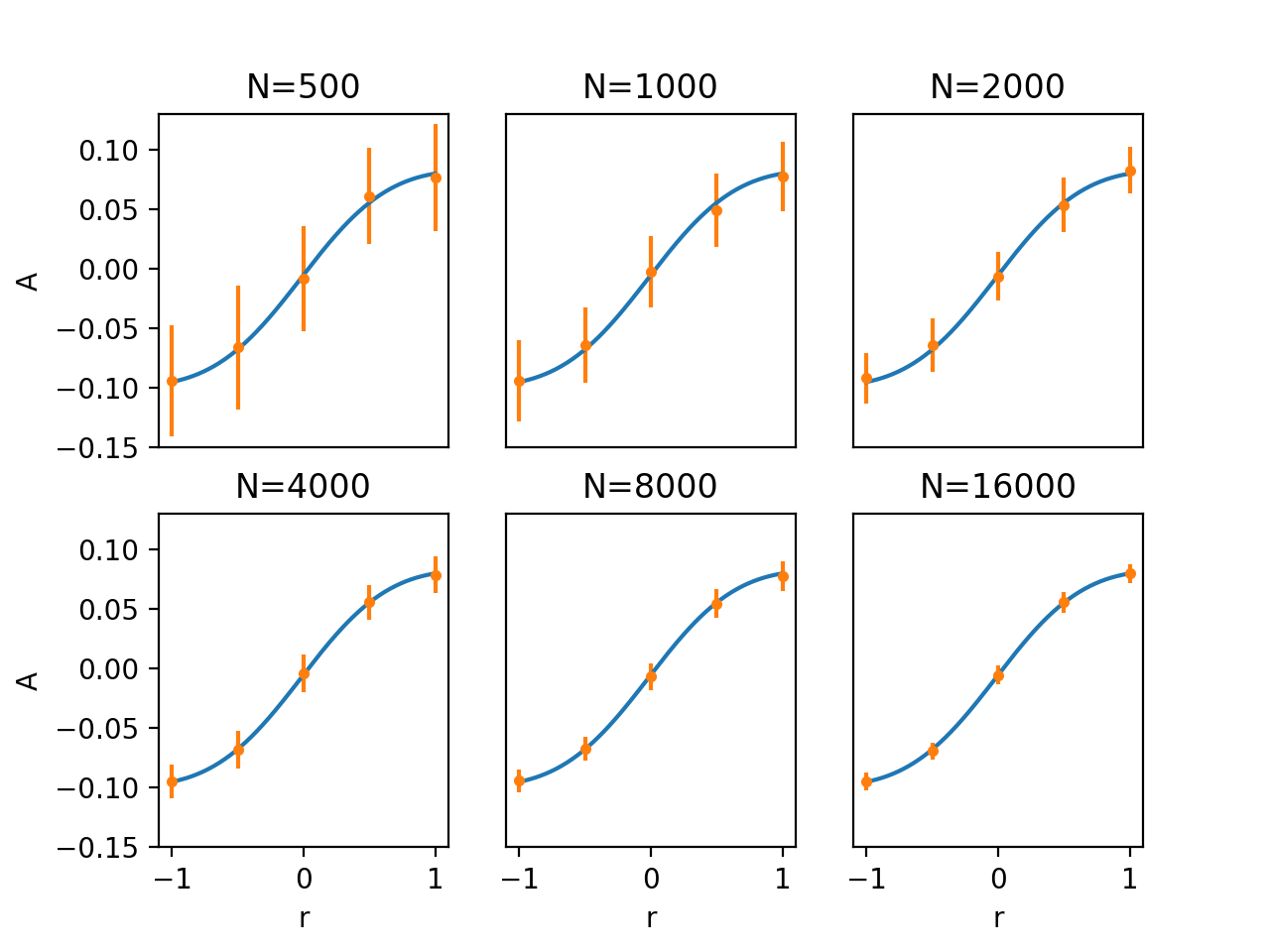}
	\caption{The fluctuation of $A$ defined in Eq. (\ref{eq:test_quantity}) in terms of $r$,
	where $N =N_1 + N_3 + N_2 + N_4$. The blue solid curve is theoretical result
	calculated from Eq.~\eqref{eq:ratio} directly. The yellow dots are the results from Monte Carlo simulation. 
	The errors reflect statistic uncertainty. The numbers of the generated events in the selected region are 500, 1000, 2000,  4000, 8000, 16000.}
	\label{fig:N500_16000}
	\end{figure}

\section{Results and Discussions}\label{sec:method}
We use Eq.~(\ref{eq:amp_square_r}) to generate the Dalitz plot distribution and Monte Carlo method
to simulate the measurement. In more detail, for a certain value of $r$, Eq.~(\ref{eq:amp_square_r})
is applied to calculate the theoretical events in region ``1", ``2", ``3", and ``4" as shown
in the right panel of Fig.~\ref{fig:interference}, i.e. $N_{1}$,
$N_{2}$, $N_{3}$ and $N_{4}$ with $N= N_{1} + N_{2} + N_{3} + N_{4}$.
We generate $N=$ 500, 1000, 2000, 4000, 8000, and 16000 events in the selected region
to check dependance of the error on the total event number $N$, as shown by Fig.~\ref{fig:N500_16000}.
Figure~\ref{fig:N500_16000} shows the dependance of $A$
defined in Eq. (\ref{eq:test_quantity}) on $r$. The value $A$ increases with
the increasing of $r$ from $-1$ to $1$. The blue solid curve is the theoretical result and the 
dots are the results from Monte Carlo method. The error bars reflect the statistic uncertainties.
The simulated total events $N_t$ in the whole phase space
depends on $N$ and $r$. Applying Eq.~~(\ref{eq:amp_square_r}), we can calculate $N_t/N$
in terms of $r$. For $r=-1,~-0.5,~0,~0.5$ and $1$,  the ratio $N_t/N$ is calculated to be $6.86,~5.61,~4.85,~5.41$
and $6.56$. We take the biggest value $6.86$ in later discussion. 

With $r$ varies from $[-1,~1]$, $A$ defined in Eq.~(\ref{eq:test_quantity}) monotonically increases from
$-0.095$ to $0.08$, which makes it efficient to tell the isospin of $T_{cc}^+$.
Assuming $T_{cc}^+$ is an eigenstate of isospin, its isospin is either $0$ or $1$, and $r$ would be either $-1$ or $1$.
From Fig. \ref{fig:N500_16000}, we can see that even a relevant small number of events, say $N=500$
and $N_t = 3430$, would be enough to distinguish the isospin of $T_{cc}^+$.
This kind of study is inspired by the study of the patterns of different isospin final states from the Dalitz plot distribution of $\eta\to\pi^+\pi^-\pi^0$ decay~\cite{Gardner:2019nid},
where different partitions of the Dalitz plot can define asymmetries sensitive to different isospin.
Analogously, we define a quantity to efficiently investigate the isospin property of $T_{cc}^+$ taking advantage of
the distribution property of the interference term as shown in Fig.~\ref{fig:interference}. The margins of the
partitions divided by $t_b$, $u_b$ and $u_c$ are chosen to maximize the discrepancies of the
quantity defined in Eq.~\eqref{eq:test_quantity} at $r=-1$ and $1$.

This formula can also apply to analyze the isospin property of other exotic candidates,
both hidden charm/bottom and double charm/bottom exotic candidates. 
For the hidden charm/bottom exotic candidates, their isospin property
can also be judged by their decay mode to charmonium and bottomonium plus several pions.
For instance, the isospin property of the $Z_c(3900)$ can be learnt as isospin triplet
due to its $J/\psi\pi$ decay channel. For open double charmonium-like/bottominium-like
exotic candidates, they do not have hidden charmonium/bottomonium decay channels. 
An alternative way is to study the property in open charm/bottom channels, i.e. 
in the open charm decay modes and open bottom decay modes, as discussed 
above for the $T_{cc}^+$. The method can also be applied to hidden charm/bottom
exotic hadrons, especially for those with large isospin violation, to analyze their isospin property.
Taking the $X(3872)$ as an example, in the $D\bar{D}^*+c.c.$ 
molecular picture,  because it extremely closes to the $D^0\bar{D}^{*0}+c.c.$
threshold, it strongly couples to this channel other than the charged one.   
The wave function for the $X(3872)$ with a given isospin reads as
\begin{eqnarray}
X(3872)=\frac 12\left[ \left( D^{*0}\bar{D}^0+D^{0}\bar{D}^{*0}\right)\pm \left(D^{*+}D^-+D^+D^{*-}\right)\right]
\end{eqnarray}
with the $+$ and $-$ signs for isospin triplet and singlet, respectively. 
Analogous to the $T_{cc}^+$, the three-body decay modes $X(3872)\to D^0\bar{D}^{*0}\to D^0D^-\pi^+$
($X(3872)\to D^{*0}\bar{D}^0\to D^+\bar{D}^0\pi^-$)
and $X(3872)\to D^{*+}D^{-}\to D^0D^-\pi^+$ ($X(3872)\to D^+D^{*-}\to D^+\bar{D}^0\pi^-$) have the same final particles and can be 
used to analyze the isospin property of the $X(3872)$.

\section{Summary}\label{sec:summary}

We introduce an efficient method to investigate the isospin property of the $T_{cc}^+$ from the Dalitz plot
distribution of the $T_{cc}^+\to D^+D^-\pi^0$ process, which happens through the two different intermediate
states, i.e. $D^{*0}D^+$ and $D^0D^{*+}$. The interference of the two diagrams
is significantly different around the intersection of the $D^{*0}$ and $D^{*+}$
signal in Dalitz plot.  Accordingly, we propose a measurable quantity $A$ which monotonically
depends on $r$, with $r$ the key value to determine the isospin property of a given particle.  
This method can also be used for other exotic candidates. For instance, the $D^0D^-\pi^+$ and $D^+\bar{D}^0\pi^-$ 
three-body modes for the $X(3872)$. This method can be used in further experimental analysis,
especially for double charm/bottom exotic candidates which do not have hidden charmonium/bottomonium decay channels. 

\section{Acknowledgement}
Discussions with Jian Liang and Jifeng Hu are acknowledged.
This work is partly supported by Guangdong Major Project of Basic and Applied Basic Research No.~2020B0301030008,
the National Natural Science Foundation of China with Grant No.~12035007,
Guangdong Provincial funding with Grant No.~2019QN01X172 and the National Natural Science Foundation of China with
Grant No.~12105108.
Q.W. is also supported by the NSFC and the Deutsche Forschungsgemeinschaft (DFG, German
Research Foundation) through the funds provided to the Sino-German Collaborative
Research Center TRR110 ``Symmetries and the Emergence of Structure in QCD"
(NSFC Grant No. 12070131001, DFG Project-ID 196253076-TRR 110).

\end{document}